\documentclass[twocolumn,twoside]{IEEEtran}
\usepackage{amsmath,amssymb,amsthm,wasysym,epsfig,dsfont,color,subfigure,empheq,graphicx}
\usepackage{enumerate,url,algpseudocode,algorithm,empheq,balance,url}
\usepackage[table]{xcolor}

\newtheorem{proposition}{Proposition}

\theoremstyle{remark}
\allowdisplaybreaks

\begin{document}
\title{Ergodic Energy Management Leveraging\\
Resource Variability in Distribution Grids}

\author{
Gang Wang,~\IEEEmembership{Student Member,~IEEE,}
	Vassilis Kekatos,~\IEEEmembership{Member,~IEEE,}\\
	Antonio J. Conejo,~\IEEEmembership{Fellow,~IEEE,} and
	Georgios B. Giannakis,~\IEEEmembership{Fellow,~IEEE}
	

\thanks{Manuscript received July 9, 2015; revised October 7, and December 22, 2015; accepted January 30, 2016. Date of publication DATE; date of current version DATE. Paper no. TPWRS.00983.2015. Work in this paper was supported by NSF grants 1423316, 1442686, 1508993, and 1509040. G. Wang and G. B. Giannakis are with the Digital Technology Center and the ECE Dept., University of Minnesota, Minneapolis, MN 55455, USA. G. Wang is also with the School of Automation, Beijing Institute of Technology, Beijing 100081, P. R. China. V. Kekatos is with the ECE Dept., Virginia Tech, Blacksburg, VA 24061, USA. A. J. Conejo is with the ISE and ECE Depts., The Ohio State University, Columbus, OH 43210, USA. Emails: \{gangwang,georgios\}@umn.edu; kekatos@vt.edu; conejonavarro.1@osu.edu.}
%
%
}


\maketitle

\begin{abstract}
Contemporary electricity distribution systems are being challenged by the variability of renewable energy sources. Slow response times and long energy management periods cannot efficiently integrate intermittent renewable generation and demand. Yet stochasticity can be judiciously coupled with system flexibilities to enhance grid operation efficiency. Voltage magnitudes for instance can transiently exceed regulation limits, while smart inverters can be overloaded over short time intervals. To implement such a mode of operation, an ergodic energy management framework is developed here. Considering a distribution grid with distributed energy sources and a feed-in tariff program, active power curtailment and reactive power compensation are formulated as a stochastic optimization problem. Tighter operational constraints are enforced in an average sense, while looser margins are enforced to be satisfied at all times. Stochastic dual subgradient solvers are developed based on exact and approximate grid models of varying complexity. Numerical tests on a real-world 56-bus distribution grid and the IEEE 123-bus test feeder relying on both grid models corroborate the advantages of the novel schemes over their deterministic alternatives.
\end{abstract}


\section{Introduction}\label{sec:intro}
Distributed generation and the prospective integration of electric vehicles and elastic loads create unseen operational scenarios for distribution grids~\cite{proc2011chertkov}. Several utilities in the US currently experience issues with integrating residential- and commercial-scale solar generation. For example, solar farms oftentimes connected at the end of a long distribution feeder in distant rural areas, are routinely reported to introduce voltage regulation problems to the residential buses across the feeder. These frequently reversing power flows strain the apparent power capabilities of substation transformers. Moreover, data collected from residential PVs reveal that solar generation can fluctuate by up to $15\%$ of their nameplate ratings within one-minute intervals~\cite{SmartStar}. The aforementioned issues critically challenge energy management of distribution grids.

On the other hand, contemporary distributed generation units are equipped with the so-termed \emph{smart power inverters} that have two-way communication and computing capabilities, and thus offer unprecedented control opportunities~\cite{2014colin}. Leveraging smart inverters for joint reactive power compensation and active power curtailment to achieve various objectives (power loss minimization, conservation voltage reduction, or voltage regulation) is considered here. Traditionally, distribution grid voltage regulation is performed via load-tap-changing transformers, capacitor banks, and voltage regulators~\cite{Kersting}. This utility-owned equipment involves discrete control actions, and its lifespan is affected by frequent switching operations~\cite{tse2015jabr}, \cite{2014colin}. Regulating voltage under increasing generation from distributed renewable sources would require even more frequent switching actions and perhaps further installations.

In this context, recent research efforts have focused on engaging smart inverters in the energy management system (EMS) of distribution grids~\cite{proc2011chertkov}, \cite{Carvalho}, \cite{tec2014chertkov}; especially, given that these inverters come with PVs and electric vehicles anyways. Customer-owned power inverters can be commanded to adjust reactive power injections within milliseconds and in a continuously-valued manner~\cite{nrel2008liub}, \cite{Carvalho}. Albeit currently prohibited by some standards (see e.g., \cite{IEEE1547}), controlling reactive power through smart inverters has been reported to improve grid's voltage profile, or even displace utility-owned voltage regulating equipment at more than $50\%$ solar penetration~\cite{nrel2008liub}.

Using approximate grid models, voltage regulation is effected through a multi-agent scheme in \cite{Markabi}, and local control algorithms are devised in \cite{tec2014chertkov}. Building on the exact full AC grid model, reactive power control is an instance of the optimal power flow (OPF) problem, which is non-convex in general~\cite{tac2015gan}. Recently, different convex relaxations have been proposed; see \cite{Low14} for a review. In radial networks, the OPF can be relaxed into a second-order cone program (SOCP) via either polar coordinates \cite{Jabr06}, or the branch flow model \cite{FL1}; or into a semidefinite program (SDP) \cite{Bai08}. Although the two relaxations have been shown to be equivalent, \cite{BoseAllerton} advocates using the SOCP one due to its lower computational complexity. Leveraging the SOCP relaxation, a two timescale conventional and inverter-based reactive power control has been formulated in~\cite{sgc2011farivar}. Accounting for stochasticity, an adaptive local control algorithm for single-branch grids is developed in~\cite{tps2012yeh}, and a stochastic centralized approach has been pursued in~\cite{tps2015vassilis}.

Apart from exploiting the reactive power capabilities of smart inverters, active power curtailment has been advocated as an ancillary service as well~\cite{tse2011tonkoski,tse2014emiliano, tse2014su,tsg2014appen}. Droop-based active power curtailment has been proposed as an efficient means for overvoltage prevention \cite{tse2011tonkoski}. In \cite{tse2014emiliano}, an SDP-based relaxation has been devised for jointly commanding active and reactive power setpoints to inverters in multi-phase distribution grids. Leveraging joint reactive power compensation and active power curtailment, a multi-objective OPF is formulated for unbalanced four-wire distribution grids in \cite{tse2014su}. Local voltage control strategies are developed for customers enrolled in a feed-in tariff (FIT) policy \cite{tsg2014appen}. An FIT power supply policy compensates DG owners for the surplus of renewable energy they inject into the grid. Similarly structured programs have been successfully deployed in Europe and several US states~\cite{fit2009}.

Existing energy management schemes enforce inverter-related and voltage regulation-related constraints at all times. However, the operation of future grids could benefit from currently unexploited system flexibilities. Two possible options are the overloading tolerance of inverters and the voltage regulation margins. Specifically, the inverters found in DG units and storage units are manufactured to operate higher than their nameplate apparent power rating \cite{book2014blackburn}. Actually, this feature has already been exploited in designing panels~\cite{gtd2013kirtley}. Moreover, most voltage regulation standards, such as the American National Standard Institute (ANSI) C84.1 \cite{ansic84}, and the EN~50160 standard \cite{en50160}, define two voltage regions: one for normal operations, and one whose use is limited in frequency and duration. According to the EN~50160 standard, for example, nodal voltage magnitudes are required to lie in the latter region for $95\%$ of any $10$-minute sample~\cite{tse2015jabr},~\cite{en50160}. 

To exploit such flexibilities, this work proposes an energy management scheme, where voltage regulation and inverter capacity constraints are imposed in a stochastic rather than a deterministic sense. Our contribution is not on the effect of pricing and curtailment policies on renewable integration. It is rather an algorithmic framework for exploiting the aforementioned sources of flexibility to lower costs and improve renewable integration. Different from existing schemes, operational constraints are relaxed instantaneously, while tighter limits are enforced in an average sense. This is achieved using a stochastic dual subgradient scheme that relies on power flow models with different accuracy-complexity trade-offs. The schemes are based only on data to command set-points to DG units, and enjoy convergence and feasibility guarantees. Numerical tests using synthetic and real data on a 56-bus and the IEEE 123-bus grids corroborate the efficacy of the scheme.

\emph{Paper outline.} The rest of the paper is outlined as follows. The novel energy management problem is formulated in Section~\ref{sec:problem}. An ergodic optimization approach is presented in Section \ref{sec:eem}, while a stochastic approximation solver is developed in Section~\ref{sec:sas}. The implementation of the solver depending on two grid models is presented in Section~\ref{sec:model}, while performance advantages over the deterministic alternatives are supported by the numerical tests of Section~\ref{sec:tests}. Concluding remarks are drawn in Section~\ref{sec:con}.

\emph{Notation.} Lower- (upper-) case boldface letters denote column vectors (matrices), with the exception of power flow vectors $(\mathbf{P},\mathbf{Q})$. Calligraphic symbols are reserved for sets, while $\mathbb{R}^{\rm N}_+$ denotes the set of all nonnegative $N$-dimensional vectors; the symbol $^{\top}$ stands for transposition; and $\mathbf{0}$ and $\mathbf{1}$ denote the all-zeros and all-ones vectors, respectively. 

\section{Problem Formulation}\label{sec:problem}
{Consider a distribution grid comprising $N+1$ buses. The grid is modeled by a tree graph $\mathcal{T}:=(\mathcal{N}_o,\mathcal{L})$, where $\mathcal{N}_o:=\{0,1,\ldots,N\}$ is the set of nodes (buses) and $|\mathcal{L}|=N$ denotes the cardinality of the edge (line) set $\mathcal{L}$. Note that albeit structurally meshed, distribution grids are usually operated as radial.} The tree is rooted at the substation bus indexed by $n=0$, and all non-root buses comprise the set $\mathcal{N}:=\{1,\ldots,N\}$. Let $v_n$ be the squared voltage magnitude at bus $n$, and $p_n+j q_n$ the complex power injection at bus $n$ for all $n\in\mathcal{N}_o$. For notational brevity, nodal quantities related to non-root buses are stacked in column vectors $\mathbf{v}$, $\mathbf{p}$, and $\mathbf{q}$.

Active and reactive power injections at bus $n$ are split into their generation and consumption components as $p_{n}:=p_{n}^g-p_{n}^c$ and $q_{n}:=q_{n}^g-q_{n}^c$. For a purely load bus, the consumption components $(p_{n}^c,q_{n}^c)$ are oftentimes related via a constant power factor, whereas $p_{n}^g=q_{n}^g=0$. A DG bus in addition to the nonnegative components $(p_n^c,q_n^c)$, it also provides renewable generation $p_{n}^g\ge 0$ and reactive power support $q_{n}^g$. All buses are henceforth assumed to be constant power buses. For buses having only a shunt capacitor, it holds that $p_{n}^g=p_{n}^c=q_{n}^c=0$ and $q_{n}^g>0$. Generation and consumption components are stacked accordingly in vectors $\mathbf{p}^c$, $\mathbf{p}^g$, $\mathbf{q}^c$, and $\mathbf{q}^g$. 

The energy management controller is run centrally by the utility and communicates set-points to DG units. Although coordinative control of power inverters and utility-owned voltage regulating devices would only improve performance, it is a nontrivial task and is not considered here; see~\cite{sgc2011farivar} for a dynamic programming approach. In the envisioned scenario, the grid operation is divided into short control periods indexed by $t$. The duration of these periods depends on the variability of active and reactive power consumption and the availability of data predictions. Since power inverters provide a continuously-valued control variable that can be adjusted in milliseconds, transients have been reported to be negligible~\cite{nrel2008liub}. This is in contrast to conventional voltage regulating equipment that results in switching transients. Without loss of generality, a 30-sec control interval will be presumed hereafter.
 
During time period $t$, the grid operator can buy or sell energy $p_{0,t}$ from or to the main grid through the substation bus via the real-time market. The price for this energy exchange is $\pi_{0,t}$, and it is assumed to be positive at all times. Apparently, if the real-time market operates on a 5-min basis, the price $\pi_{0,t}$ remains constant over 10 consecutive control periods. Internally in the distribution grid, customers with renewable generation units, such as PVs or wind micro-turbines, can subscribe to a so-termed FIT program; see e.g.,~\cite{fit2009}, \cite{2013fit}. According to this program, the surplus of renewable energy a customer may have can be bought if deemed appropriate by the utility company at the FIT price $\pi_{f,t}$. Although FIT prices are currently adjusted on a monthly basis, time-varying prices $\pi_{f,t}$ are considered here for the sake of generality. Feed-in-tariff prices are also assumed to be positive. Energy consumption from both FIT and regular customers is charged at a retail price $\pi_{r,t}$. The energy cost for customer $n$ during period $t$ is the product of 
\begin{equation}\label{eq:customer-cost}
\pi_{r,t}[p_{n,t}^g-p_{n,t}^c]_{-} -\pi_{f,t}[p_{n,t}^g-p_{n,t}^c]_+
\end{equation}
times the duration of the control period, where the operators $[a]_+:=\max\{a,0\}$ and $[a]_{-}:=\max\{0,-a\}$ \cite{tse2013zhang}. Apparently, at most one of the two summands in \eqref{eq:customer-cost} is nonzero per slot $t$.

{From the utility side, the energy cost for time slot $t$ is
\begin{equation}\label{eq:utility-cost}
\pi_{0,t}\,p_{0,t}(\mathbf{p}_t^g,\mathbf{q}_t^g)+\pi_{f,t}\mathbf{1}^\top[\mathbf{p}_t^g-\mathbf{p}_t^c]_{+} - \pi_{r,t}\mathbf{1}^\top[\mathbf{p}_t^g-\mathbf{p}_t^c]_{-}
\end{equation}
multiplied by the slot duration, where $[\mathbf{a}]_{+}$ and $[\mathbf{a}]_{-}$ for vectors are applied entry-wise now.} Heed that the energy exchange with the main grid $p_{0,t}(\mathbf{p}_t^g,\mathbf{q}_t^g)$ depends on the internal consumption and generation, and the associated power losses on distribution lines. Thus, the energy exchange $p_{0,t}$ can be thought of as a function of the control variables $(\mathbf{p}_t^g,\mathbf{q}_t^g)$, while its dependence on $(\mathbf{p}_t^c,\mathbf{q}_t^c)$ and grid power losses is implicitly indicated by the subscript $t$.

If the energy management scheme were to minimize the utility's cost in \eqref{eq:utility-cost}, it would force the minimum possible local generation. To see that, consider a node $n$ where at time $t$ the demand is higher than the installed solar capacity; hence, $-[p_{n,t}^g-p_{n,t}^c]_{-}=p_{n,t}^g-p_{n,t}^c<0$. Then, the utility EMS would command $p_{n,t}^g=0$ unless there is an under-voltage condition. Such a policy contradicts the purpose of an FIT program. The FIT program should curtail renewable power only if a customer has a surplus and the surplus cannot be bought due to either financial or voltage regulation reasons. To accommodate the FIT terms, the utility does not curtail renewable generation when the net injection is negative. Thus, the cost to be minimized by the energy management scheme is $\pi_{0,t}\,p_{0,t}(\mathbf{p}_t^g,\mathbf{q}_t^g)+\pi_{f,t}\mathbf{1}^\top[\mathbf{p}_t^g-\mathbf{p}_t^c]_{+}$ rather than that in \eqref{eq:utility-cost}.

Operation of the energy management scheme is detailed next. Before control period $t$, the EMS collects predictions for prices $(\pi_{0,t},\pi_{f,t})$, loads $(\mathbf{p}^c_t,\mathbf{q}_t^c)$, and the maximum renewable generation $\overline{\mathbf{p}}_t^g$. At every period $t$, buses are partitioned to those having a renewable energy surplus comprising the set
\begin{equation}\label{eq:Nt}
\mathcal{N}_t:=\{n\in \mathcal{N}:~\overline{p}_{n,t}^g\geq p_{n,t}^c\}
\end{equation}
and to those buses belonging to the complementary set of $\mathcal{N}_t$ denoted by $\overline{\mathcal{N}}_t$. To jointly perform active power curtailment and reactive power management, the EMS could solve the ensuing problem per time interval $t$
\begin{subequations}\label{eq:trad}
\begin{align}
J_{1,t}^\ast:=\min_{\mathbf{p}_t^g,\mathbf{q}_t^g}~&~\pi_{0,t}\,p_{0,t}(\mathbf{p}_t^g,\mathbf{q}_t^g)+\pi_{f,t}\mathbf{1}^\top[\mathbf{p}_t^g-\mathbf{p}_t^c]_{+}\label{eq:trad1}\\
\textrm{s.to}~&~ 0\le p_{n,t}^g\le \overline{p}_{n,t}^g,~\forall ~n\in\mathcal{N}_t\label{eq:trad2a}\\
~&~ p_{n,t}^g=\overline{p}_{n,t}^g,~\forall ~n\in\overline{\mathcal{N}}_t\label{eq:trad2b}\\
~&~(p_{n,t}^g)^2+(q_{n,t}^g)^2\le s_n^2,~\forall~n\label{eq:trad3}\\
~&~v_l\leq v_{n,t}(\mathbf{p}_t^g,\mathbf{q}_t^g) \leq v_u,~\forall~n.\label{eq:trad4}
\end{align}
\end{subequations}
Power injections $\{(p_{n,t}^g,q_{n,t}^g)\}_n$ are constrained in the feasible set defined by \eqref{eq:trad2a}-\eqref{eq:trad4}. Constraints \eqref{eq:trad2a}-\eqref{eq:trad3} apply locally per bus $n$, whereas the voltage constraints in \eqref{eq:trad4} couple power injections across the grid. Specifically, the term $\overline{p}_{n,t}^g-p_{n,t}^g$ in \eqref{eq:trad2a} is the active power curtailed for all inverters with renewable surplus at time $t$, i.e., $n\in\mathcal{N}_t$. Constraint \eqref{eq:trad3} originates from the maximum apparent power capability (nameplate rating) $s_n$ of inverter $n$. Constraint \eqref{eq:trad4} maintains the squared voltage magnitudes in the prescribed interval $\mathcal{V}:=[v_l,~v_u]$ at all nodes. Similar to the energy exchange $p_{0,t}(\mathbf{p}_t^g,\mathbf{q}_t^g)$, voltage magnitudes are expressed as implicit functions of $(\mathbf{p}_t^g,\mathbf{q}_t^g)$, whose actual function forms depend on the postulated grid model and are elaborated in Section~\ref{sec:model}. To simplify the exposition, constraints on the apparent power flows on distribution lines have been ignored; such limits can be readily incorporated using the grid model of Section~\ref{subsec:ldf}. It is worth mentioning that policy scenarios where the utility accepts any energy surplus as soon as grid constraints are satisfied can be captured by simply setting FIT prices $\pi_{f,t}$ to zero for all $t$ in \eqref{eq:trad}.

Problem \eqref{eq:trad} guarantees that all power and voltage constraints are satisfied at all times. Nevertheless, future distribution grids will afford flexibilities that can be leveraged to lower operational costs and better integrate renewables. Two possible sources of flexibility are the overloading capability of smart inverters and the voltage regulation ranges. Regarding the former, a grid-tied power inverter can exceed its apparent power capacity for a short period of time.  Indeed, power electronics are empirically designed to operate at even 1.2-1.5 times higher than their nameplate rating~\cite{book2014blackburn}. For the latter, instead of requiring the squared voltage magnitudes to lie in $\mathcal{V}$ at every $t$, it suffices for their time-averages to lie in $\mathcal{V}$, and the instantaneous values to lie within a wider range. For instance, according to standard EN 50160, voltages are required to stay in $\mathcal{V}$ for $95\%$ of any $10$-minute sample~\cite{en50160}. Additionally, heed that problem \eqref{eq:trad} depends on predictions $(\mathbf{p}_t^c$, $\mathbf{q}_t^c$, $\overline{\mathbf{p}}_t^g)$, and prices $(\pi_{0,t}$, $\pi_{f,t})$. It is therefore optimal only if load demand, renewable generation, and prices are perfectly known. In practice though $\left(\mathbf{p}_t^c, \mathbf{q}_t^c,\overline{\mathbf{p}}_t^g,\pi_{0,t},\pi_{f,t}\right)$ involve uncertainties (e.g. measurement noise, time-delay, and system variability) rendering the solution of \eqref{eq:trad} hardly optimal.

To leverage operational flexibilities and cope with uncertainties, a stochastic rather than the deterministic energy management formulation of \eqref{eq:trad} is pursued next. {To that end, the time-varying problem parameters $\left\{\mathbf{p}_t^c,\mathbf{q}_t^c,\overline{\mathbf{p}}_t^g,\pi_{0,t},\pi_{f,t}\right\}$ are modeled as stochastic processes~\cite{rser2009carta}, \cite{jors2006conejo}, \cite{SPM2013}. To capture ensemble averages via time averages, the aforementioned stochastic processes are assumed stationary and ergodic, yet not necessarily independent across time; see \cite{book2010conejo} and \cite{icassp2014gatsis}. Recall that a stochastic process is ergodic if its moments (e.g., the mean) can be inferred from a single realization of the process.} The novel energy management scheme entails solving the following stochastic optimization problem
\begin{subequations}\label{eq:eem}
\begin{align} 
J_2^\ast{:=}\min_{\left\{\mathbf{p}_t^g,\mathbf{q}_t^g\right\}_t}&~\mathbb{E}\left[\pi_{0,t}p_{0,t}(\mathbf{p}_t^g,\mathbf{q}_t^g)+\pi_{f,t}\mathbf{1}^\top[\mathbf{p}_t^g-\!\mathbf{p}_t^c]_{+}\right]\label{eq:eem:cost}\\
\textrm{s.to}~~&~0\le p_{n,t}^g\le \overline{p}_{n,t}^g,~\forall~n \in\mathcal{N}_t\label{eq:eem:solar}\\
~&~p_{n,t}^g=\overline{p}_{n,t}^g,~\forall ~n\in\overline{\mathcal{N}}_t\label{eq:eem:nocurtail}\\
~&~(p_{n,t}^g)^2+(q_{n,t}^g)^2\le \overline{s}_n^2,~\forall~n\label{eq:eem:apc}\\
~&~\underline{v}_l\leq v_{n,t}(\mathbf{p}_t^g,\mathbf{q}_t^g)\leq \overline{v}_u,~\forall~n\label{eq:eem:vrc}\\~
~&~\mathbb{E}\left[(p_{n,t}^g)^2+(q_{n,t}^g)^2\right]\le s_n^2,~\forall~n\label{eq:eem:aapc}\\
~&~v_l\leq \mathbb{E}\left[v_{n,t}(\mathbf{p}_t^g,\mathbf{q}_t^g)\right] \leq v_u,~\forall~n\label{eq:eem:avrc}
\end{align}
\end{subequations}
where the optimization variables consist of $(\mathbf{p}_t^g,\mathbf{q}_t^g)$ for all periods $t$, and the expectations are taken over the joint distribution of $\left(\mathbf{p}_t^c,\mathbf{q}_t^c,\overline{\mathbf{p}}_t^g,\pi_{0,t},\pi_{f,t}\right)$ across all periods $t$.
Constraint \eqref{eq:eem:aapc} guarantees that the \emph{average} apparent power complies with the nameplate inverter capacity for all buses; while constraint \eqref{eq:eem:apc} enforces a hard limit $\overline{s}_n$ $(\overline{s}_n\geq s_n)$ on the instantaneous apparent power for all $n$. Similarly, the averages of squared voltage magnitudes are maintained in $\mathcal{V}$ according to \eqref{eq:eem:avrc}, whereas constraint \eqref{eq:eem:vrc} ensures that their instantaneous values lie in a region $\mathcal{V}':=[\underline{v}_l,~\overline{v}_u]$ with $\mathcal{V}\subseteq \mathcal{V}'$. For example, the ANSI C84.1 requires voltage magnitudes to lie within $\mathcal{V}=[0.95^2,1.05^2]$~per unit (p.u.) of normal operation, but within  $\mathcal{V}'=[0.917^2,1.058^2]$~p.u. over short durations~\cite{ansic84}.

Let us compare the solution of \eqref{eq:eem} to the minimizers obtained from \eqref{eq:trad} at every time $t$. Note that constraint \eqref{eq:trad3} implies constraints \eqref{eq:eem:apc} and \eqref{eq:eem:aapc}, but not the converse. Likewise, constraints \eqref{eq:trad4} guarantees \eqref{eq:eem:vrc} and \eqref{eq:eem:avrc}. Therefore, the stochastic scheme in \eqref{eq:eem} constitutes a relaxation of the deterministic problem \eqref{eq:trad} solved over time $t$. As such, the minimizers of \eqref{eq:eem} could yield a lower \emph{average} operational cost, i.e., $J_2^\ast\le \mathbb{E}[J_{1,t}^\ast]$, where the expectation is taken over time $t$.

The stochastic problem in \eqref{eq:eem} involves infinitely many variables $\{\mathbf{p}^g_t,\mathbf{q}^g_t\}_t$. Nodal power injections at time $t$ should satisfy the instantaneous constraints \eqref{eq:eem:solar}--\eqref{eq:eem:vrc}. Further, the infinitely many variables are coupled across time via the objective function and the average constraints \eqref{eq:eem:aapc} and \eqref{eq:eem:avrc}, hence challenging the solution of \eqref{eq:eem}. A stochastic optimization approach for tackling \eqref{eq:eem} is pursued in the next section.

\section{Ergodic Energy Management}\label{sec:eem}
The goal of ergodic energy management (EEM) is to design algorithms that sequentially observe predictions $\left(\mathbf{p}_t^c,\mathbf{q}_t^c,\overline{\mathbf{p}}_t^g,\pi_{0,t},\pi_{f,t}\right)$, and solve near optimally the stochastic problem in \eqref{eq:eem}. The EEM is inspired by related ideas from resource allocation in wireless communication networks, where due to propagation channel uncertainties and variabilities, one prefers to optimize the average rather than the instantaneous system behavior~\cite{twc2010ngargg},~\cite{alegg}. The key assumption is that only realizations of those stochastic processes are available, while their joint probability density function is typically unknown.

Since optimization variables, henceforth collectively denoted by $\mathbf{x}:=\left(\{\mathbf{p}^g_t,\mathbf{q}^g_t\}_t\right)$, are coupled via expectations, constraints \eqref{eq:eem:aapc} and \eqref{eq:eem:avrc} are dualized. Let $\boldsymbol{\nu}\in\mathbb{R}_+^{\rm{N}}$, $\underline{\boldsymbol{\xi}}\in\mathbb{R}_+^{\rm{N}}$, and $\overline{\boldsymbol{\xi}}\in\mathbb{R}_+^{\rm{N}}$ denote the dual variables corresponding to \eqref{eq:eem:aapc}, and the lower and upper voltage bounds in \eqref{eq:eem:avrc}, respectively. All other constraints are kept explicit. Using these definitions, the Lagrangian function of \eqref{eq:eem} is readily written as
\begin{align}
\mathcal{L}\left(\mathbf{x};\boldsymbol{\nu},\underline{\boldsymbol{\xi}},\overline{\boldsymbol{\xi}}\right)
 :=&~\mathbb{E}\left[\pi_{0,t}\,p_{0,t}(\mathbf{p}_t^g,\mathbf{q}_t^g)+\pi_{f,t}\mathbf{1}^\top[\mathbf{p}_t^g-\mathbf{p}_t^c]_{+}\right]\nonumber\\
&+\sum_{n=1}^N\nu_n\left\{\mathbb{E}\left[(p_{n,t}^g)^2+(q_{n,t}^g)^2\right]-s_n^2\right\}\nonumber\\
&+\sum_{n=1}^N\underline{\xi}_n\left\{v_l-\mathbb{E}\left[v_{n,t}(\mathbf{p}_t^g,\mathbf{q}_t^g)\right]\right\}\nonumber\\
&+ \sum_{n=1}^N\overline{\xi}_n\left\{\mathbb{E}\left[v_{n,t}(\mathbf{p}_t^g,\mathbf{q}_t^g)\right]- v_u\right\}\label{eq:lag}.
\end{align}

The dual function for problem \eqref{eq:eem} is the minimum of the Lagrangian function over all primal variables. Due to the linearity of the expectation operator, the minimization and the expectation operators can be interchanged. After rearranging terms, the dual function is thus expressed as
\begin{align*}
g(\boldsymbol{\nu},\underline{\boldsymbol{\xi}},\overline{\boldsymbol{\xi}}):=\mathbb{E}\left[g_t(\boldsymbol{\nu},\underline{\boldsymbol{\xi}},\overline{\boldsymbol{\xi}})\right] -\sum_{n=1}^N\left(\nu_n{s}_n^2-\underline{\xi}_n v_l + \overline{\xi}_n v_u\right)
\end{align*}
where functions $g_t(\boldsymbol{\nu},\underline{\boldsymbol{\xi}},\overline{\boldsymbol{\xi}})$ are defined as
\begin{align}\label{eq:sep}
g_t(\boldsymbol{\nu},\underline{\boldsymbol{\xi}},\overline{\boldsymbol{\xi}}):= \min_{(\mathbf{p}_t^g,\mathbf{q}_t^g)\in\Omega_t}&\Big\{
\pi_{0,t}p_{0,t}(\mathbf{p}_t^g,\mathbf{q}_t^g)+\pi_{f,t}\mathbf{1}^\top[\mathbf{p}_t^g-\mathbf{p}_t^c]_{+}\nonumber\\
&+\sum_{n=1}^N\nu_n\left[(p_{n,t}^g)^2+(q_{n,t}^g)^2\right] \nonumber\\
&+\sum_{n=1}^N(\overline{\xi}_n-\underline{\xi}_n)v_{n,t}(\mathbf{p}_t^g,\mathbf{q}_t^g)\Big\}
\end{align} 
and the feasible set $\Omega_t$ is given by the instantaneous constraints in \eqref{eq:eem} as
\begin{equation}\label{eq:Omega}
\Omega_t:=\left\{(\mathbf{p}_t^g,\mathbf{q}_t^g)~\textrm{satisfying}~\eqref{eq:eem:solar}-\eqref{eq:eem:vrc}\right\}.
\end{equation}

The dual problem is obtained by maximizing the dual function over the dual variables, that is
\begin{align}\label{eq:dualval}
g(\boldsymbol{\nu}^\ast,\underline{\boldsymbol{\xi}}^\ast,\overline{\boldsymbol{\xi}}^\ast):=\max_{\boldsymbol{\nu},\underline{\boldsymbol{\xi}},\overline{\boldsymbol{\xi}}\ge\mathbf{0}} g(\boldsymbol{\nu},\underline{\boldsymbol{\xi}},\overline{\boldsymbol{\xi}}).
\end{align}
Evaluating $g(\boldsymbol{\nu},\underline{\boldsymbol{\xi}},\overline{\boldsymbol{\xi}})$ requires solving infinitely many problems of the form in \eqref{eq:sep}, and then averaging the optimal costs over the joint probability density function (pdf) of $\left\{\mathbf{p}_t^c,\mathbf{q}_t^c,\mathbf{\overline{p}}_t^g,\pi_{0,t},\pi_{f,t}\right\}$. Even if the joint pdf were available, finding the expectations would be non-trivial. Hence, even evaluating the dual function becomes challenging. To maximize the dual function in a feasible manner, a stochastic optimization solver is proposed next.

\section{Stochastic Approximation Solver}\label{sec:sas}
The problem at hand is tackled using a stochastic dual subgradient method~\cite{twc2010ngargg,alegg,tps1999conejo}. To maximize $g(\boldsymbol{\nu},\underline{\boldsymbol{\xi}},\overline{\boldsymbol{\xi}})$, the Lagrange multipliers are updated using the projected subgradient iterations for some step size $\mu>0$, as
\begin{subequations}\label{eq:dualvar}
\begin{align}
{\boldsymbol{\nu}}_t &:=\left[{\boldsymbol{\nu}}_{t-1}+\mu\boldsymbol{\delta}_{\nu,t}\right]_+\label{eq:updatenu}\\
{\underline{\boldsymbol{\xi}}}_t &: =\left[{\underline{\boldsymbol{\xi}}}_{t-1}+\mu\boldsymbol{\delta}_{\underline{\xi},t}\right]_+\label{eq:updatexil}\\
{\overline{\boldsymbol{\xi}}}_t &: =\left[{\overline{\boldsymbol{\xi}}}_{t-1}+\mu\boldsymbol{\delta}_{\overline{\xi},t}\right]_+\label{eq:updatexiu}
\end{align} 
\end{subequations}
where the vector $\boldsymbol{\delta}_t:=[\boldsymbol{\delta}_{\nu,t}^{\top}~ \boldsymbol{\delta}_{\underline{\xi},t}^{\top}~\boldsymbol{\delta}_{\overline{\xi},t}^{\top}]^{\top}$ is a subgradient of $g_t(\boldsymbol{\nu},\underline{\boldsymbol{\xi}},\overline{\boldsymbol{\xi}})$ evaluated at the previous iterate $({\boldsymbol{\nu}}_{t-1},{\underline{\boldsymbol{\xi}}}_{t-1},{\overline{\boldsymbol{\xi}}}_{t-1})$. The entries of the subgradient vector, denoted by $[\boldsymbol{\delta}_{t}]_n$, can be found as 
\begin{subequations}\label{eq:dualvarupdates}
\begin{align}
[\boldsymbol{\delta}_{\nu,t}]_n&:= (\hat{p}_{n,t}^g)^2+(\hat{q}_{n,t}^g)^2-s_{n}^2\label{eq:subgnu}\\
[\boldsymbol{\delta}_{\underline{\xi},t}]_n&:= v_l-v_{n,t}(\hat{\mathbf{p}}_t^g,\hat{\mathbf{q}}_t^g)\label{eq:subksilower}\\
[\boldsymbol{\delta}_{\overline{\xi},t}]_n&:= v_{n,t}(\hat{\mathbf{p}}_t^g,\hat{\mathbf{q}}_t^g)-v_u\label{eq:subgksiupper}
\end{align}
\end{subequations}
for all $n$, where $(\hat{\mathbf{p}}_t^g,\hat{\mathbf{q}}_t^g)$ are the minimizers of the problem in \eqref{eq:sep} for $g_t({\boldsymbol{\nu}}_{t-1},{\underline{\boldsymbol{\xi}}}_{t-1},{\overline{\boldsymbol{\xi}}}_{t-1})$. Note that the Lagrange multipliers are updated at every control interval.

\begin{table}[t]
\renewcommand{\arraystretch}{1.2}
\caption{Ergodic Energy Management Algorithm}\label{tbl:eem}
\small
\vspace{-0.5em}
\begin{tabular}{|p{0.9\linewidth}|}
\hline
\vspace{-.3em}
1: Input operational limits $\{s_n,\overline{s}_n\}_{n\in\mathcal{N}}$, $({v}_l,{v}_u)$, $(\underline{v}_l,\overline{v}_u)$, and step size $\mu>0$.\\
2: Dual variables ${\boldsymbol{\nu}}_{0}$, ${\underline{\boldsymbol{\xi}}}_{0}$, and ${\overline{\boldsymbol{\xi}}}_{0}$ are initialized to zero.\\
3: \textbf{For} $t=1,2,\ldots$ \textbf{do}\\
4:\hspace*{.5em} Acquire predictions $(\mathbf{p}^c_t,\mathbf{q}_t^c,\overline{\mathbf{p}}_t^g,\pi_{0,t},\pi_{f,t})$.\\
5:\hspace*{.5em} Find primal variables $(\hat{\mathbf{p}}_t^g,\hat{\mathbf{q}}_t^g)$ as the minimizers of $g_t({\boldsymbol{\nu}}_{t-1},{\underline{\boldsymbol{\xi}}}_{t-1},{\overline{\boldsymbol{\xi}}}_{t-1})$ by solving \eqref{eq:primalbf} or \eqref{eq:primalldf}.\\
6:\hspace*{.5em} Update dual variables $({\boldsymbol{\nu}}_t,{\underline{\boldsymbol{\xi}}}_t,{\overline{\boldsymbol{\xi}}}_t)$ using \eqref{eq:dualvar}.\\
7:\hspace*{.5em} Communicate setpoints $(\hat{\mathbf{p}}_t^g,\hat{\mathbf{q}}_t^g)$ to DGs.\\
8: \textbf{End for}\\
\hline
\end{tabular}
\vspace{-0.5em}
\color{black}
\end{table}

Table \ref{tbl:eem} summarizes the EEM algorithm. Operational limits as well as the step size are set in Step 1, and Lagrange multipliers are initialized to zero in Step 2. The EEM then iterates between four steps. In Step 4, the utility collects predictions for the random variables involved. In the absence of more elaborate options, the most recently observed or metered values can be used as predictions for the upcoming control period of interest. Step 5 finds the optimal primal variables by solving \eqref{eq:sep} evaluated at the current value of the Lagrange multipliers. Step 6 updates the Lagrange multipliers via the dual subgradient rule of \eqref{eq:dualvar}. The calculated setpoints $(\hat{\mathbf{p}}_t^g,\hat{\mathbf{q}}_t^g)$ are finally communicated to the DGs, and applied to the grid in Step 7. It is worth stressing that the proposed EEM scheme does not require any distributional knowledge on the input data $\left(\mathbf{p}_t^c,\mathbf{q}_t^c,\overline{\mathbf{p}}_t^g,\pi_{0,t},\pi_{f,t}\right)$. Moreover, although the focus is on utility cost minimization, other energy management tasks such as voltage regulation and conservation voltage reduction, could be cast under this framework.

As far as convergence is concerned, note first that all primal and dual iterates depend on the realizations $(\mathbf{p}_t^c,\mathbf{q}_t^c,\overline{\mathbf{p}}_t^g,\pi_{0,t},\pi_{f,t})$, and are thus random. For that reason, convergence claims are in probability. 
Using the definition of $\boldsymbol{\delta}_t$, 
it is easy to show that there exists a finite $H$ such that $\mathbb{E}\big[\|\boldsymbol{\delta}_t\|_2^2\big|{\boldsymbol{\nu}}_{t-1},{\underline{\boldsymbol{\xi}}}_{t-1},{\overline{\boldsymbol{\xi}}}_{t-1}\big]\le {H}^2$ for all $t$, i.e., the subgradient $\boldsymbol{\delta}_t$ is bounded at all times. In particular, it holds that $[\boldsymbol{\delta}_{\nu,t}]_n^2\leq \overline{s}_n^2$, while $[\boldsymbol{\delta}_{\underline{\xi},t}]_n^2$ and $[\boldsymbol{\delta}_{\underline{\xi},t}]_n^2$ are both upper bounded by $(\overline{v}^u-\underline{v}_l)^2$. Thus, the bound $H$ can be selected as 
\begin{equation}\label{eq:H}
H:=\sum_{n=1}^N \left[\overline{s}_n^2+2(\overline{v}^u-\underline{v}_l)^2\right].
\end{equation}
Adopting \cite[Theorem~1]{alegg}, the following result characterizes the almost sure feasibility and optimality of the EEM algorithm.
\begin{proposition}[\cite{alegg}]\label{pro:rate}
For the sequences $\{\hat{\mathbf{p}}_t^g,\hat{\mathbf{q}}_t^g\}_t$ generated by the algorithm in Table~\ref{tbl:eem}, the next hold with probability $1$ for all $n\in\mathcal{N}$
\begin{subequations}\label{eq:asfeas}
\begin{align}
&\lim_{t\to\infty}\frac{1}{t}\sum_{\tau=1}^t[(\hat{p}_{n,\tau}^g)^2+(\hat{q}_{n,\tau}^g)^2]\le s_n^2 \label{eq:asfeas3}\\
&~v_l\le\lim_{t\to\infty}\frac{1}{t}\sum_{\tau=1}^t v_{n,\tau}(\hat{\mathbf{p}}_\tau^g,\hat{\mathbf{q}}_\tau^g)\le v_u \label{eq:asfeas4}.
\end{align}
\end{subequations}
Furthermore, the incurred operational costs satisfy
\begin{equation*}
\lim_{t\to\infty}\frac{1}{t}\sum_{\tau=1}^t\Big[ \pi_{0,\tau}\,p_{0,\tau}(\hat{\mathbf{p}}_{\tau}^g,\hat{\mathbf{q}}_{\tau}^g)+\pi_{f,\tau}\mathbf{1}^\top [\hat{\mathbf{p}}_{\tau}^g-\mathbf{p}_\tau^c]_{+}\Big]-J_2^\ast\le\frac{\mu H^2}{2}
 \end{equation*} 
almost surely for $H$ as in \eqref{eq:H}.
\end{proposition}
The proof of Proposition~\ref{pro:rate} can be found in~\cite{alegg}. Proposition~\ref{pro:rate} asserts that the ensembles of primal sequences $\{\hat{\mathbf{p}}_t^g,\hat{\mathbf{q}}_t^g\}_t$ are feasible almost surely, meaning that constraints \eqref{eq:eem:aapc} and \eqref{eq:eem:avrc} are satisfied almost surely. Moreover, the ergodic limit of the objective is at most $\mu H^2/2$ away from the optimal $J_2^\ast$ [cf. \eqref{eq:eem}]. The aforementioned claims hold even if the stochastic processes involved are correlated across time~\cite{alegg}. Although stochastic processes have been assumed to be ergodic for the theoretical claims to hold, the numerical tests in Section~\ref{sec:tests} using real data show the efficacy of the scheme even with non-ergodic data. 

The EEM problem in \eqref{eq:eem} and its stochastic approximation solver of Table~\ref{tbl:eem} involve the power losses $p_{0,t}(\mathbf{p}_t^g,\mathbf{q}_t^g)$ and the squared voltage magnitudes $\{v_{n,t}(\mathbf{p}_t^g,\mathbf{q}_t^g)\}_n$. So far, both quantities have been expressed as functions of the control variables $(\mathbf{p}_t^g,\mathbf{q}_t^g)$. In that respect, the EEM scheme constitutes a general framework where different power system models can be assumed. To implement Step 5 in the algorithm of Table~\ref{tbl:eem}, the actual forms of $p_{0,t}(\mathbf{p}_t^g,\mathbf{q}_t^g)$ and $\{v_{n,t}(\mathbf{p}_t^g,\mathbf{q}_t^g)\}_n$ need to be specified. As a turnkey application of EEM, the ensuing section focuses on radial single-phase distribution grids using two power flow models with different accuracy-complexity trade-offs.
\color{black}

\section{Grid Modeling and Algorithms}\label{sec:model}

This section specifies functions $p_{0,t}(\mathbf{p}_t^g,\mathbf{q}_t^g)$ and $\{v_{n,t}(\mathbf{p}_t^g,\mathbf{q}_t^g)\}_n$ using an exact full AC grid model and its linear approximation. Both cases are then integrated into the EEM algorithm. Selecting between the two models relies on the computational capabilities that can be afforded. The AC model-based EEM can be formulated as an SOCP, whereas the linear model yields a linearly constrained quadratic program. Therefore, the latter option offers an approximate yet computationally less demanding alternative to the former.


\subsection{Branch Flow Model-based EEM}\label{subsec:bfm}
\begin{figure}[t]
\centering
\includegraphics[scale=0.3]{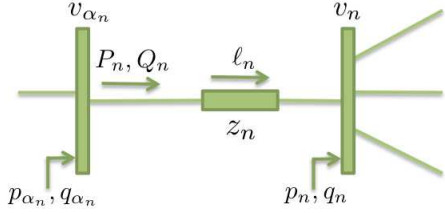}
\caption{Bus $n$ is connected to its unique parent $\alpha_n$ via line $n$.}
\label{fig:diagram}
\end{figure}

Due to the radial structure of distribution grids, every non-root bus $n\in\mathcal{N}$ has a unique parent bus, which will be denoted by $\alpha_n$. The directed edge $(\alpha_n,n)\in\mathcal{L}$ corresponding to the distribution line feeding bus $n$ will be indexed by $n$; see Fig.~\ref{fig:diagram}. Without loss of generality, buses can be indexed such that $\alpha_n < n$ for all $n\in\mathcal{N}$. 

Let $z_n=r_n+jx_n$ be the line impedance of line $n$, and $\ell_{n,t}$ the squared current magnitude on line $n$ at time $t$. If $P_{n,t}+jQ_{n,t}$ is the complex power flow on line $n$ seen at the sending end bus $\alpha_n$ at time $t$, the grid can be described by the branch-flow model~\cite{tpd1989baranwu}
\begin{subequations}\label{eq:model}
\begin{align}
p_{n,t}&=\sum_{k\in\mathcal{C}_n}P_{k,t}  - (P_{n,t} -r_n \ell_{n,t})\label{eq:mp}\\
q_{n,t}&=\sum_{k\in\mathcal{C}_n}Q_{k,t}  - (Q_{n,t} -x_n \ell_{n,t})\label{eq:mq}\\
v_{n,t}&=v_{\alpha_n,t}- 2(r_nP_{n,t}+x_nQ_{n,t}) +(r_n^2+x_n^2)\ell_{n,t} \label{eq:mv}\\
\ell_{n,t}&=\frac{P_{n,t}^2+Q_{n,t}^2}{v_{\alpha_n,t}}\label{eq:ml}
\end{align}
\end{subequations}
for all $n\in \mathcal{N}$, where $\mathcal{C}_n$ is the set of the children nodes of bus $n$. The power injections at the substation bus are $p_{0,t}=\sum_{k\in\mathcal{C}_0}P_{k,t}$, $q_{0,t}=\sum_{k \in \mathcal{C}_0} Q_{k,t}$, and its squared voltage magnitude $v_{0,t}$ is controlled at a desirable value. Similar to $(\mathbf{p}_t,\mathbf{q}_t)$, the vectors $\mathbf{r}$, $\mathbf{P}_t$, $\mathbf{Q}_t$, $\mathbf{v}_t$, and $\boldsymbol{\ell}_t$, collect the entries of $r_n$, $P_{n,t}$, $Q_{n,t}$, $v_{n,t}$, and $\ell_{n,t}$, accordingly.

Equations \eqref{eq:mp}-\eqref{eq:mv} are linear with respect to the system variables $(\mathbf{p}_t,\mathbf{q}_t,\mathbf{P}_t,\mathbf{Q}_t,\mathbf{v}_t,\boldsymbol{\ell}_t)$. The equations in \eqref{eq:ml} are quadratic yielding a non-convex feasible set. Nonetheless, the latter equations have been recently relaxed to convex inequalities described by the hyperbolic constraints~\cite{FL1} 
\begin{align}\label{eq:inequality}
P_{n,t}^2+Q_{n,t}^2\le v_{\alpha_n,t}\ell_{n,t},~\forall~n
\end{align}
which can be equivalently expressed as the convex second-order cone constraints
\begin{align}\label{eq:soc}
\left\|\left[\begin{array}{c}
2P_{n,t}\\
2Q_{n,t}\\
v_{\alpha_n,t}-\ell_{n,t}\end{array}\right]\right\|_2\le v_{\alpha_n,t}+\ell_{n,t},~\forall~n.
\end{align}
Equations \eqref{eq:mp}-\eqref{eq:mv} and \eqref{eq:soc} represent now a convex feasible set. Recent works suggest using this relaxed feasible set to perform several grid optimization tasks. Under different conditions, the relaxation has been shown to be exact, i.e., the obtained minimizer attains \eqref{eq:soc} with equality; see~\cite{tac2015gan} and references therein. Henceforth, all SOCP relaxations are assumed exact, which will be numerically verified in Section~\ref{sec:tests}. 

Based on \eqref{eq:mp}-\eqref{eq:mv} and \eqref{eq:soc}, the active power injection at the substation bus $p_{0,t}(\mathbf{p}_t^g,\mathbf{q}_t^g)$ can be expressed as 
\begin{align*}
p_{0,t}(\mathbf{p}_t^g,\mathbf{q}_t^g)&=\sum_{n=1}^N(p_{n,t}^c-p_{n,t}^g)+\sum_{n=1}^N r_n\ell_{n,t}\\
&=\mathbf{1}^\top (\mathbf{p}_{t}^c-\mathbf{p}_{t}^g)+\mathbf{r}^\top\boldsymbol{\ell}_t
\end{align*}
where the second summand represents the total power losses on distribution lines. Hence, under the aforementioned relaxed grid model, the primal update (Step 5 in Table \ref{tbl:eem}) entails solving the optimization problem
\begin{align}\label{eq:primalbf}
\min_{\mathbf{p}_t^g,\mathbf{q}_t^g, \boldsymbol{\ell}_t,\atop\mathbf{P}_t,\mathbf{Q}_t,\mathbf{v}_t}&~
\pi_{0,t}\mathbf{1}^\top (\mathbf{p}_{t}^c-\mathbf{p}_{t}^g)+\pi_{0,t}\mathbf{r}^\top\boldsymbol{\ell}_t
+\pi_{f,t}\mathbf{1}^\top[\mathbf{p}_t^g-\mathbf{p}_t^c]_{+}\nonumber\\
&+
\sum_{n=1}^N{\nu}_{n,t-1}\left[(p_{n,t}^g)^2+(q_{n,t}^g)^2\right]
\nonumber\\
&+
\sum_{n=1}^N({\overline{\xi}}_{n,t-1}-{\underline{\xi}}_{n,t-1})v_{n,t}\\
\textrm{s.to}~&~\eqref{eq:eem:solar}-\eqref{eq:eem:vrc},\eqref{eq:mp}-\eqref{eq:mv}, \eqref{eq:soc}
\nonumber.
\end{align}

In addition to the original variables $(\mathbf{p}_t^g,\mathbf{q}_t^g)$, the primal update now involves the variables $(\mathbf{P}_t,\mathbf{Q}_t,\mathbf{v}_t,\boldsymbol{\ell}_t)$ too. Problem \eqref{eq:primalbf} can be reformulated to an SOCP. All instances of \eqref{eq:primalbf} solved in Section \ref{sec:tests} were exact. Nevertheless, solving \eqref{eq:primalbf} could be computationally demanding for large-scale distribution grids. This motivates our next instantiation of the EEM algorithm under an approximate grid model. 

\subsection{Linear Distribution Flow-Based EEM}\label{subsec:ldf}
The linear distribution flow model can be derived as follows. Because the line parameters $\{r_n,x_n\}_{n\in\mathcal{N}}$ have relatively small entries, the last summands in \eqref{eq:mp}-\eqref{eq:mv} can be ignored yielding the linear equations for all $n\in\mathcal{N}$~\cite{tpd1989baranwu}
\begin{subequations}\label{eq:linear}
\begin{align}
p_{n,t}&=\sum_{k\in\mathcal{C}_n}P_{k,t}  - P_{n,t}\label{eq:mcp}\\
q_{n,t}&=\sum_{k\in\mathcal{C}_n}Q_{k,t}  - Q_{n,t} \label{eq:mcq}\\
v_{n,t}&=v_{\alpha_n,t}- 2(r_nP_{n,t}+x_nQ_{n,t}).\label{eq:mcv}
\end{align}
\end{subequations}
In this way, squared voltage magnitudes are now approximated as linear functions of $(\mathbf{p}_t,\mathbf{q}_t)$. {Assuming squared voltage magnitudes to be close to 1 p.u.,} squared line current magnitudes are approximated as~\cite{tpd1989baranwu}
\begin{equation}\label{eq:iqol}
\ell_{n,t}=\frac{P_{n,t}^2+Q_{n,t}^2}{v_{\alpha_n,t}}\approx P_{n,t}^2+Q_{n,t}^2.
\end{equation}
Therefore, the active power injection at the substation bus can be thus approximated by
\begin{equation*}
p_{0,t}(\mathbf{p}_t^g,\mathbf{q}_t^g)=\mathbf{1}^\top(\mathbf{p}_t^c-\mathbf{p}_t^g)+\sum_{n=1}^N r_n\left(P_{n,t}^2+Q_{n,t}^2\right).
\end{equation*}

Building on the approximate model of \eqref{eq:linear}--\eqref{eq:iqol}, the primal update of the EEM algorithm (Step 5 of Table~\ref{tbl:eem}) for period $t$ entails solving the problem
\begin{align}\label{eq:primalldf}
\min_{\mathbf{p}_t^g,\mathbf{q}_t^g,\atop\mathbf{P}_t,\mathbf{Q}_t,\mathbf{v}_t} 
~&~\pi_{0,t}\mathbf{1}^\top(\mathbf{p}_t^c-\mathbf{p}_t^g)+\pi_{0,t}\sum_{n=1}^N r_n(P_{n,t}^2+Q_{n,t}^2)
\nonumber\\
~&~+\pi_{f,t}\mathbf{1}^\top[\mathbf{p}_t^g-\mathbf{p}_t^c]_{+}+\sum_{n=1}^N\!({\overline{\xi}}_{n,t-1}-{\underline{\xi}}_{n,t-1})v_{n,t}\nonumber\\
~&~+\sum_{n=1}^N{\nu}_{n,t-1} \left[(p_{n,t}^g)^2+(q_{n,t}^g)^2\right]\\
\textrm{s.to}~&~ \eqref{eq:eem:solar}-\eqref{eq:eem:vrc},\eqref{eq:mcp}-\eqref{eq:mcv}.\nonumber
\end{align}
From \eqref{eq:mcp}-\eqref{eq:mcq}, the line flow variables $(\mathbf{P}_t,\mathbf{Q}_t)$ can be substituted as linear functions of $(\mathbf{p}_t,\mathbf{q}_t)$. Hence, problem \eqref{eq:primalldf} can be solved as a linearly constrained quadratic program.

\section{Numerical Tests}\label{sec:tests}
The novel schemes were numerically tested on a 56-bus distribution grid from Southern California Edison (SCE) and the IEEE 123-bus feeder~\cite{tac2015gan},~\cite{distfeeder}. Line and bus data for the SCE grid are listed in Tables~\ref{tbl:linedata} and \ref{tbl:busdata}, accordingly, while a power factor of 0.8 is assumed for all loads; see {\cite{tac2015gan} for details}. The capacity of the PVs installed on buses 19 and 45 was set to 6 MW. At each 30-sec control period, the EEM controller collects power demands from load buses and solar generation predictions from PV units. Subsequently, active and reactive power injections by PV inverters are determined by: i) solving the deterministic energy management (DEM) scheme in \eqref{eq:trad}; and ii) the novel EEM algorithm of Table~\ref{tbl:eem} that is initialized to zero. 

\begin{table}
{{
\renewcommand{\arraystretch}{1}
\caption{Line data for the 56-bus distribution feeder~\cite{tac2015gan}}\vspace*{-0.5em}
\label{tbl:linedata} \centering
\begin{tabular}{|cccc|cccc|}
\hline
From\!&\! To & $r_i$&$x_i$ & From & To & $r_i$ & $x_i$ \\
Bus\,\!&\!Bus\, & $[\Omega]$ & $[\Omega]$ & Bus\, & Bus\, & $[\Omega]$ & $[\Omega]$ \\
\hline
1&2&0.160&0.388        &28&29&0.395&0.369	\\
2&3&0.824&0.315        &29&30&0.248&0.232	\\
2&4&0.144&0.349        &30&31&0.279&0.260	\\
4&5&1.026&0.421        &32&33&0.263&0.073	\\
4&6&0.741&0.466       &32&34&0.071&0.171\\
4&7&0.528&0.468       &34&35&0.625&0.273\\
4&20&0.138&0.334      &34&36&0.510&0.209\\
7&8&0.358&0.314		&34&38&1.062&0.406	\\
8&9&2.032&0.798	&34&41&0.115&0.278 \\
8&10&0.502&0.441		 &36&37&2.018&0.819 \\
10&11&0.372&0.327		&38&39&0.610&0.238	\\
11&12&1.431&0.999		&39&40&2.349&0.964\\
11&13&0.429&0.377		&41&42&0.159&0.384\\
13&14&0.671&0.257		&41&47&0.157&0.379\\
13&15&0.457&0.401		&42&43&0.934&0.383\\
15&16&1.008&0.385		&42&44&0.506&0.163\\
15&17&0.153&0.134		&42&45&0.095&0.195\\
17&18&0.971&0.722      &42&46&01.915&0.769\\
18&19&1.885&0.721     &47&48&1.641&0.670\\
20&21&0.251&0.096      &47&49&0.081&0.196\\
20&23&0.225&0.542      &49&50&1.727&0.709\\
21&22&1.818&0.695      &49&51&0.112&0.270\\
23&24&0.127&0.542      &51&52&0.674&0.275\\
23&25&0.284&0.687      &51&53&0.070&0.170\\
25&26&0.171&0.414		&53&54&2.041&0.780\\
26&27&0.414&0.386		&53&55&0.813&0.334\\
26&32&0.205&0.495      &53&56&0.141&0.340\\
27&28&0.210&0.196      &&&&\\
\hline
\end{tabular}
}}
\end{table}

\begin{table}
{\renewcommand{\arraystretch}{1}
\caption{Bus data for the 56-bus distribution feeder~\cite{tac2015gan}}\vspace*{-.5em}
\label{tbl:busdata} \centering
\begin{tabular}{|cc|cc|cc|}
\hline
\multicolumn{4}{|c|}{\bf Load Data}&\multicolumn{2}{c|}{\bf Load Data}\\
\hline
Bus &Peak&Bus &Peak&Bus &Peak\\
No. &[MVA]& No.&[MVA]& No. &[MVA] \\
 \cline{1-6}
3&30&			 25&0.20&  		43&1.34\\
5&0.67&			27&0.13& 		44&0.13 \\
6&0.45&			28&0.13&	46&0.67	\\
7&0.89&			29&0.07&	47&0.13	\\
8&0.07&			31&0.13&	48&0.45\\
9&0.67&			32&0.27&	50&0.20\\\cline{5-6}
10&0.45&		33&0.20&	\multicolumn{2}{c|}{\bf Shunt Capacitors}\\\cline{5-6}
11&2.23&		34&0.27&	Bus&Nameplate Capacity\\
12&0.45&		35&0.45&	No. &[Mvar]\\	\cline{5-6}
14&0.20&		36&1.34&	 19&0.6 \\
16&0.13&		37&0.13&	21&0.6\\
17&0.13&		38&0.67&	  30 &0.6\\
18&0.20&   		39&0.13&	 53&0.6\\\cline{5-6}
19&0.45&		40&0.45& 		\multicolumn{2}{c|}{\bf Base Information}\\\cline{5-6}
33&2.23&	     41&0.20& 	\multicolumn{2}{c|}{$V_{\rm base}=12$kV} \\
24&0.45&	        42&0.45&       	\multicolumn{2}{c|}{$S_{\rm base}=1$MVA}	\\     
\hline
\end{tabular}}
\end{table}

The margins for squared voltage magnitudes are set as $[v_l,~v_u]=[0.9604,1.0404]$ p.u. and $[\underline{v}_l,~\overline{v}_u]=[0.9409,1.0609]$ p.u., with nominal voltage $v_0=1$ p.u. The apparent power capability for smart inverters is set to $1.3$ times the nameplate capacity of the associated PV. Performance is evaluated in terms of the energy management cost and the instantaneous counterpart of the cost in~\eqref{eq:eem}. All algorithms were implemented using MATLAB and CVX on an Intel CPU @ $3.4$ GHz ($32$ GB RAM) computer~\cite{2008cvx}. Every run for the full AC and the linear approximation model-based algorithms on the 56-bus grid was completed in 1.5 and 1.3 seconds, respectively. The related times for the IEEE 123-bus feeder increased to 4.5 and 3 seconds, respectively. It is worth mentioning that all SOCP relaxations encountered in the ensuing experiments were feasible and exact.

\begin{figure}[t]
\centering
\includegraphics[scale=0.6]{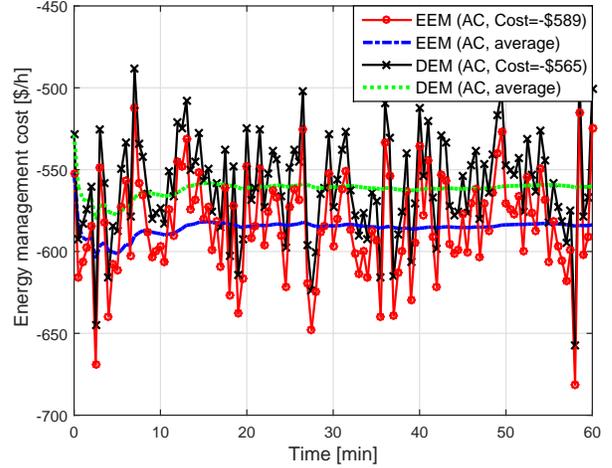}
\caption{Energy management cost using the AC model on synthetic data ($\mu=0.1$ for EEM).}
\label{fig:56cost}
\end{figure}

 \begin{figure}[t]
\centering
\includegraphics[scale=0.6]{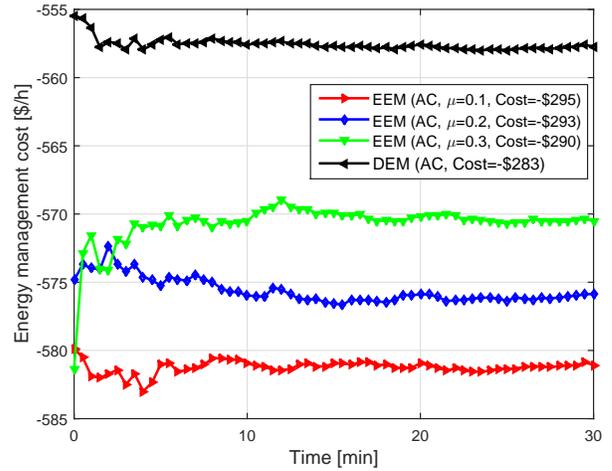}
\caption{Energy management cost averaged over 20 independent realizations using the AC model.}
\label{fig:step}
\end{figure}
 
To verify the almost sure optimality, the first experiment on the 56-bus grid simulates synthetic load consumption and solar generation as $\mathbf{p}_t^c= \mathbf{p}^c+ \boldsymbol{\epsilon}_t^c$ and $\bar{\mathbf{p}}_t^g=\mathbf{p}^g+\boldsymbol{\epsilon}_t^g$, respectively. The nominal values $\mathbf{p}^c$ and $\mathbf{p}^g$ are set to $40\%$ of the peak demand values and $80\%$ of the maximum PV generation, accordingly. Vectors $\boldsymbol{\epsilon}_t^c$ and $\boldsymbol{\epsilon}_t^g$ capture fluctuations modeled as independent zero-mean Gaussian vectors with standard deviations equal to $5\%$ of the corresponding nominal values. Given that current FIT prices change on a monthly basis and are oftentimes half of consumption prices~\cite{tsg2014appen}, prices were set to $\pi_{0,t}=30\cent/$kWh and $\pi_{f,t}=15\cent/$kWh for all $t$.

Using the branch flow model, Fig.~\ref{fig:56cost} depicts the energy management cost for the deterministic and the ergodic energy management schemes over a single realization of 120 control periods. The step size for the ergodic scheme is set to $\mu=0.1$, while the time-average energy management cost per time slot $t$ is defined as $\frac{1}{t}\sum_{\tau=1}^t\big[ \pi_{0,\tau}\,p_{0,\tau}(\hat{\mathbf{p}}_{\tau}^g,\hat{\mathbf{q}}_{\tau}^g)+\pi_{f,\tau}\mathbf{1}^\top [\hat{\mathbf{p}}_{\tau}^g-\mathbf{p}_\tau^c]_{+}\big]$. {The actual total operational cost over an hour is $-\$565$ for the DEM and $-\$589$ for the EEM scheme.}

The second test studies the effect of the step size $\mu$ on the convergence of EEM. The AC-based EEM scheme was simulated for $\mu\in\{0.1,0.2,0.3\}$ along with the DEM scheme. Twenty Monte Carlo system realizations over 60 control periods were averaged for each step size value, while the corresponding average energy management cost is plotted in Fig.~\ref{fig:step}. The curves demonstrate that larger step sizes incur higher energy management costs, an observation that agrees with the optimality gap of Proposition~\ref{pro:rate}. 

To test the proposed schemes in real-world conditions, the ensuing two experiments entail real data from the Smart$^*$ project~\cite{SmartStar}. Consumption data involved the electricity usage at minute-level samples from 443 homes on April 2, 2011; and the power output of 3 residential PVs collected at 5-second intervals over August 12, 2011. Data were preprocessed as follows. Consumption data were first linearly interpolated to yield 30-sec loads, and then averaged over every 10 homes to better resemble bus loads. Daily load curves were subsequently normalized to a maximum value of one, and mapped to different buses~\cite{tac2015gan}. Normalized daily load curves were multiplied with the nominal load value per bus. Concerning PVs, 5-sec data were aggregated to 30-sec data. Daily generation data were likewise scaled to match rated capacities.

\begin{figure}[t]
\centering
\includegraphics[scale=0.6]{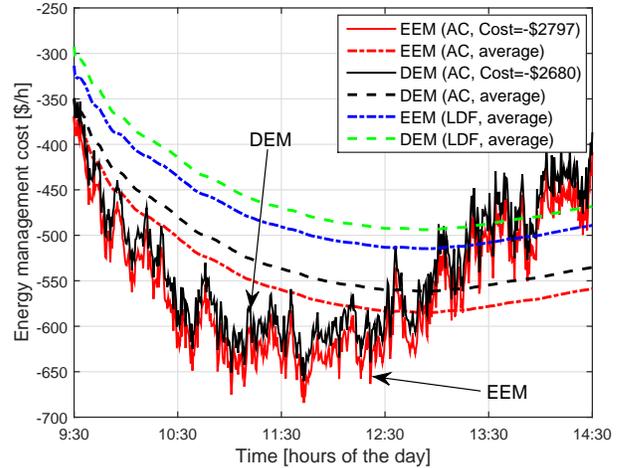}
\caption{Energy management cost using the AC and the linear distribution flow (LDF) models on real load and solar generation data~\cite{SmartStar}.}
\label{fig:track}
\end{figure}

A single system realization was simulated over the 600 30-sec control periods during 9:30am--2:30pm for both the AC- and the approximate model-based schemes. Figure~\ref{fig:track} presents the cost for $\mu=0.25$. Using either the AC or the linear approximation model, the novel EEM scheme achieves a lower cost than the DEM one.

\begin{figure}[t]
\centering
\includegraphics[scale=0.6]{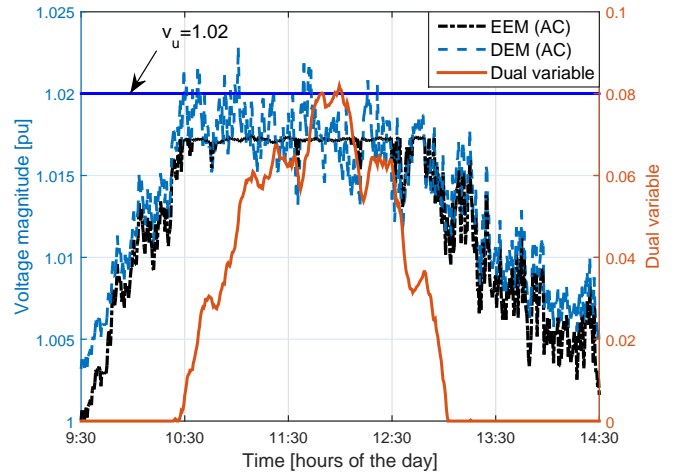} 
\caption{Voltage magnitude for bus 45 and the associated dual variable $\bar{\xi}_{45,t}$ using the AC model-based schemes.}
\label{fig:v45}
\end{figure}

\begin{figure}[htb]
\centering
\includegraphics[scale=0.6]{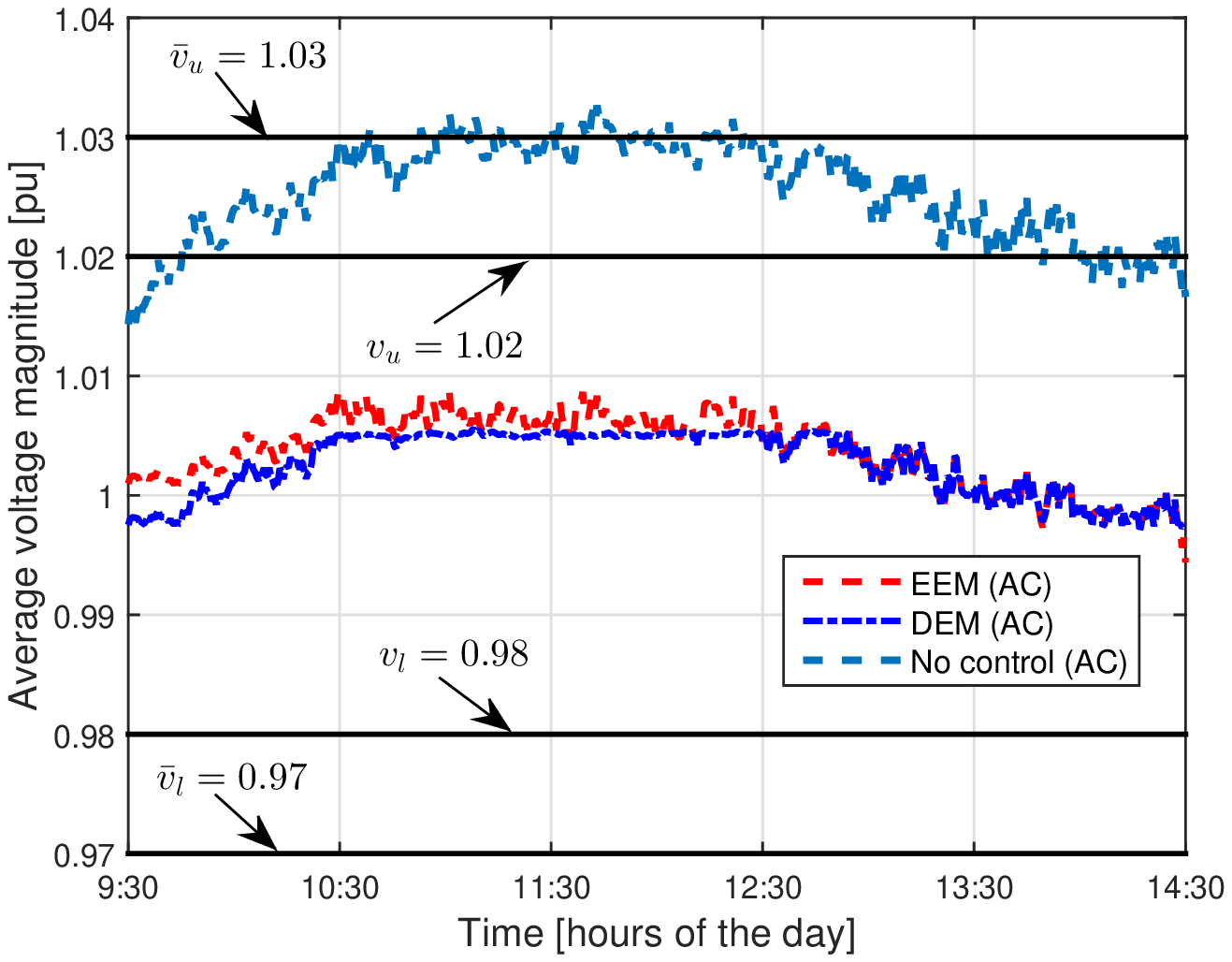} \vspace{-2.45em}\\
\includegraphics[scale=0.6]{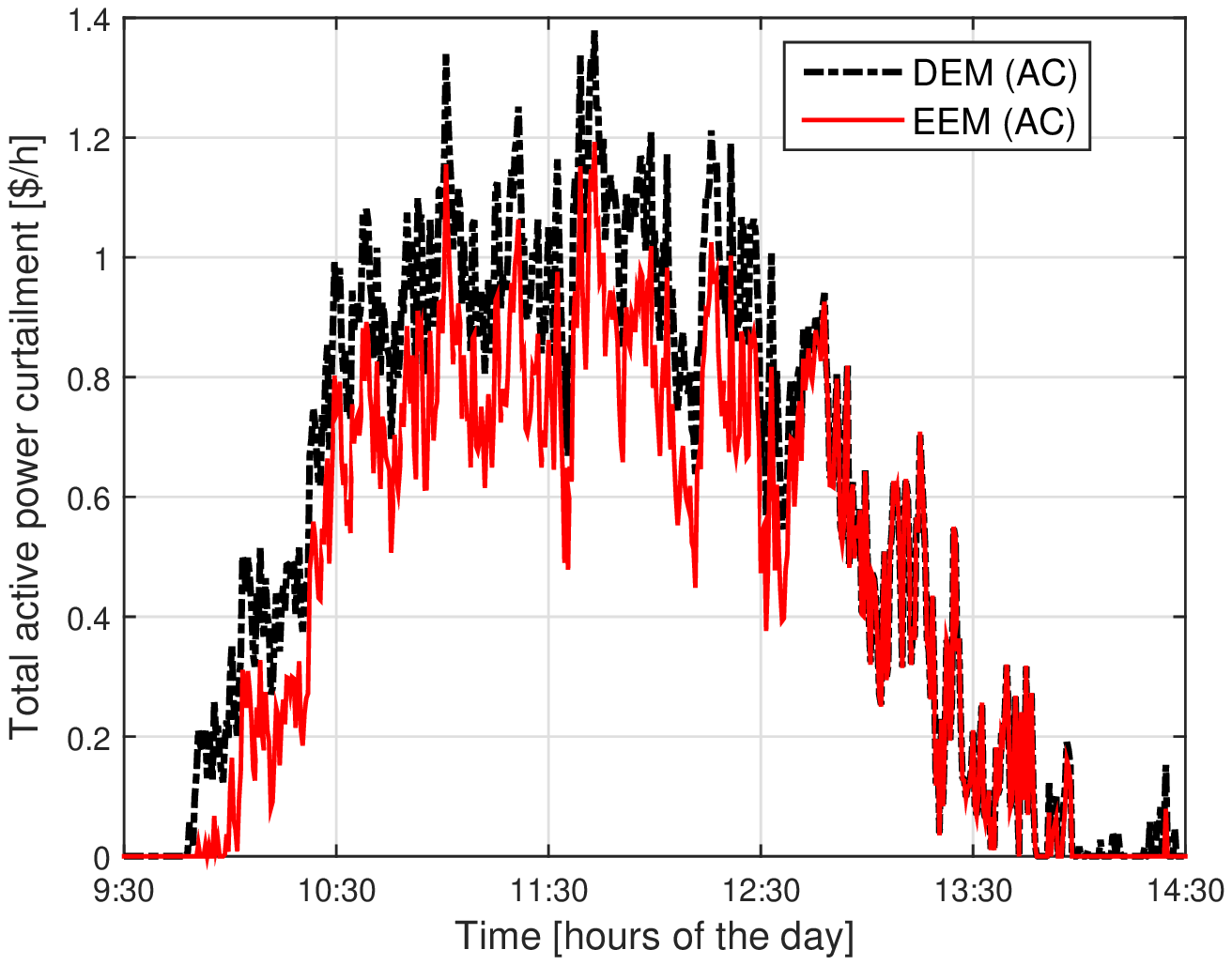} 
\caption{Top: Grid-averaged voltage magnitude using the AC model-based schemes. Bottom: Total active power curtailment over 9:30am--2:30pm.}
\label{fig:av}
\end{figure}


Fig.~\ref{fig:v45} depicts the evolution of the squared voltage magnitude for bus 45, and the evolution of the dual variable $\bar{\xi}_{45,t}$ for the tight voltage margin constraint in \eqref{eq:eem:avrc}. The voltage magnitude for the deterministic scheme remains in the tight region $[v_l,~v_u]=[0.9604,1.0404]$ throughout the operation horizon. The voltage magnitude obtained from the stochastic scheme lies occasionally beyond the voltage margin $v_u=1.0404$. Nonetheless, over-voltage effects have short duration. At around 10:25 am, when the voltage magnitude violates the tight voltage constraint for the first time, the dual variable becomes positive and starts increasing. As long as the voltage magnitude fluctuates above the tight margin, the dual variable keeps increasing. After roughly 12:20 pm, the voltage magnitude drops and remains consistently below the upper margin, while the dual variable decreases and eventually becomes zero for the rest of the day.

To get a grid-level view of voltage regulation and active power curtailment, the top panel of Fig.~\ref{fig:av} shows the grid-averaged voltage magnitude obtained via the DEM and EEM schemes, as well as without any control. Under no control, voltage magnitudes consistently exceed regulation margins. Moreover, the EEM scheme yields slightly higher voltage profile than DEM in exchange for lower operational cost. Similarly, the bottom panel of Fig.~\ref{fig:av} shows the grid-wise solar generation curtailment incurred by DEM and EEM. Apparently, the DEM scheme curtails more active power than the EEM scheme.

\begin{figure}[t]
\centering
\includegraphics[scale=0.34]{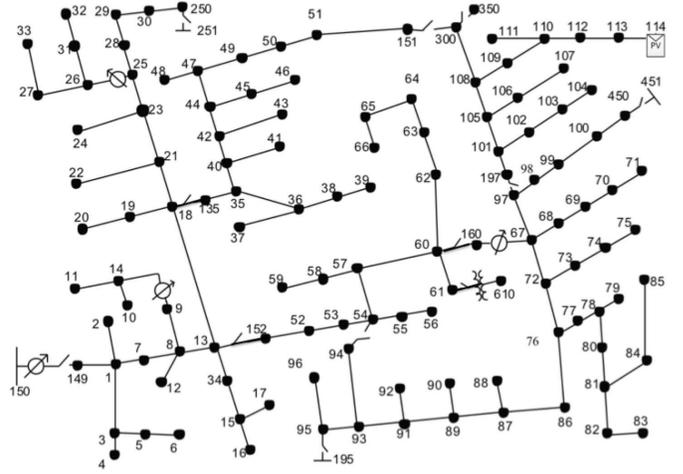}
\caption{Schematic diagram of the IEEE 123-bus test system with a PV~\cite{distfeeder}.}
\label{fig:123grid}
\end{figure}

\begin{figure}[t]
\centering
\includegraphics[scale=0.6]{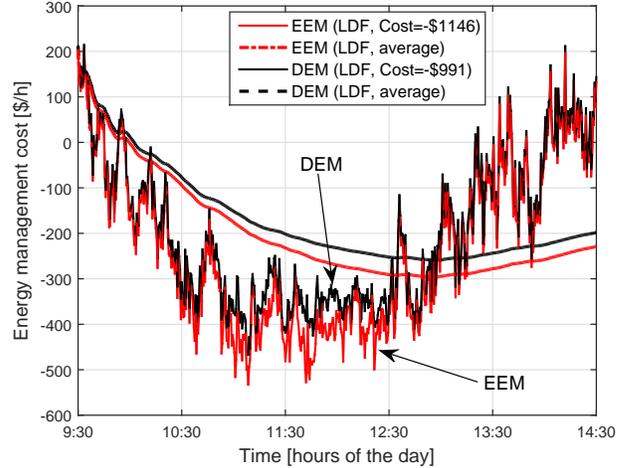}
\caption{Energy management cost evaluated on the IEEE 123-bus test system using the linearized model on real load and solar generation data.}
\label{fig:123cost}
\end{figure}

The last test involved the IEEE 123-bus feeder shown in Fig.~\ref{fig:123grid}~\cite{distfeeder}. The original multiphase system was heuristically modified to a single-phase one as described in~\cite{tac2015gan}: Loads were split uniformly over all phases. Line self-impedances were averaged over phases, while mutual impedances were neglected. Closed switches were modeled as short circuits and open switches were ignored. Distributed loads were replaced by two identical spot loads at the two line ends. Transformers were modeled as lines with given impedances, and tap ratios for all voltage regulators were fixed to 1.08. A single PV with nameplate rating of $1.2$~MW is installed at bus $114$, which corresponds to PV penetration of about 100$\%$. With $v_0=1$ p.u., voltage regulation bounds are chosen as $[v_l,~v_u]=[0.9801,1.0201]$ p.u. and $[\underline{v}_l,~\overline{v}_u]=[0.9409,1.0609]$ p.u., while inverters can be overloaded by $110$\% their nameplate rating. The linearized model was adopted, and real data for solar generation and home loads were utilized. Each time period $t$, the prices were set to $\pi_{0,t}=30\cent/$kWh and $\pi_{f,t}=15\cent/$kWh. Fig. \ref{fig:123cost} presents the cost over 600 30-sec control slots during 9:30 am -- 2:30 pm for the two schemes. The step size was set to $\mu=0.001$. The total operational cost over the simulation period amounts to $-\$1,146$ for EEM and $-\$991$ for DEM, thus demonstrating the superiority of the ergodic approach in the IEEE 123-bus feeder.

\section{Concluding Remarks}\label{sec:con}
This paper introduced an EEM framework. Smart inverters are engaged in active power curtailment and reactive power support in a stochastic sense. A stochastic dual subgradient scheme enforces tighter operational margins at all times, yet letting system characteristics deviate over short time intervals. The developed algorithms are guaranteed to converge to the optimal operational point, while the feasibility is satisfied almost surely. Numerical tests using a full AC grid model and its linear approximation on a 56-bus grid and the IEEE 123-bus feeder demonstrated the viability of the approach. In particular, the grid was operated within the regulated margins at all times, while local variables could fluctuate over looser ranges during extreme conditions. The suggested flexible grid operation brings up several interesting questions. Enforcing probabilistic rather than average constraints is worth investigating. Decentralized and localized implementations are timely and pertinent. Integrating utility-owned voltage regulating equipment to develop coordinative control schemes constitutes an interesting and challenging future research direction.

\bibliographystyle{IEEEtran}
\end{document}